\documentclass[3p,times]{elsarticle}

\usepackage{ecrc}

\volume{00}
\firstpage{1}

\journalname{Nuclear Physics A}
\runauth{Ralf-Arno Tripolt et al.}
\jid{nupha}
\jnltitlelogo{Nuclear Physics A}

\usepackage{graphicx}
\usepackage{amsmath,amssymb}
\usepackage{lineno,hyperref}
\usepackage{color}
\usepackage{tabularx}

\newcolumntype{L}[1]{>{\raggedright\arraybackslash}p{#1}} 
\newcolumntype{C}[1]{>{\centering\arraybackslash}p{#1}} 
\newcolumntype{R}[1]{>{\raggedleft\arraybackslash}p{#1}} 
\newcommand{\beq}{\begin{equation}}
\newcommand{\eeq}{\end{equation}}
\newcommand{\bea}{\begin{eqnarray}}
\newcommand{\eea}{\end{eqnarray}}

\newcommand{\pkt}{\;.}

\graphicspath{{.}{figures/}}

\biboptions{}

\begin{document}

\begin{frontmatter}

\title{Spectral functions from the functional renormalization group}

\author[TUD]{Ralf-Arno Tripolt}
\author[RKU]{Nils Strodthoff}
\author[TUD,JLU]{Lorenz von Smekal}
\author[TUD,GSI]{Jochen Wambach}

\address[TUD]{Institut f\"ur Kernphysik - Theoriezentrum,
Technische Universit\"at Darmstadt, Germany} 
\address[RKU]{Institut f\"ur Theoretische Physik,
Ruprecht-Karls-Universit\"at Heidelberg, Germany} 
\address[JLU]{Institut f\"ur Theoretische Physik,
Justus-Liebig-Universit\"at Giessen, Germany} 
\address[GSI]{GSI Helmholtzzentrum f\"ur 
Schwerionenforschung GmbH, Germany} 

\begin{abstract}
We present a viable method to obtain real-time quantities such as
spectral functions or transport coefficients at finite temperature
and density within a non-perturbative Functional Renormalization Group
approach. Our method is based on a thermodynamically consistent
truncation of the flow equations for 2-point functions with
analytically continued frequency components in the originally
Euclidean external momenta.  We demonstrate its feasibility by
calculating the mesonic spectral functions in the quark-meson model at
different temperatures and quark chemical potentials,  
in particular around the critical endpoint in the phase
diagram of the model. 
\end{abstract}


\end{frontmatter}

\section{Introduction}
\label{intro}

The calculation of real-time observables like spectral functions represents a
great challenge, due to the analytic continuation problem, to all 
Euclidean approaches to thermal Quantum Field Theory (QFT).
When based on numerical data at discrete Matsubara frequencies the
reconstruction of real-time correlations classifies as an ill-posed
inverse problem, as for example in Lattice QCD 
where techniques like the maximum entropy method (MEM) have to be used
\cite{Jarrell:1996,Asakawa:2000tr}.  
Therefore any approach that can deal with the analytic 
continuation explicitly is highly desirable. 

In the following we present such a method \cite{Kamikado2013a,
  Tripolt2014} to obtain spectral functions from the non-perturbative
Functional Renormalization Group (FRG) 
\cite{Berges:2000ew,Pawlowski:2005xe,Schaefer:2006sr,Braun:2011pp,Gies2012}
at finite temperature and density where the analytic continuation is performed 
on the level of the flow equations, as an alternative to the approach in \cite{Floerchinger2012}. 
In this way we have access to retarded propagators and real-time
spectral functions without need for any numerical reconstruction method.

Apart from being comparatively simple our method is also
thermodynamically consistent, i.e.\ the screening masses obtained from
the thermodynamic grand potential agree with those extracted from the
propagators in the space-like zero momentum limit \cite{Strodthoff:2011tz}.   
Moreover, our method satisfies the physical Baym-Mermin boundary
conditions \cite{Baym1961,Landsman1987} 
and it can be extended to include the full momentum dependence
of the 2-point functions in a computation of the grand potential
beyond leading order derivative expansion by iteration. Finally, it
can also be applied to calculate quark and gluonic spectral functions
as an alternative to analytically continued Dyson-Schwinger equations (DSEs)
\cite{Strauss:2012dg}, or to using MEM on Euclidean FRG
\cite{Haas:2013hpa} or DSE results
\cite{Nickel:2006mm,Mueller:2010ah,Qin:2013ufa}. 

As an extension to previous results for the $O(4)$ linear-sigma 
model in the vacuum \cite{Kamikado2013a}, in these proceedings, which
are based on \cite{Tripolt2014}, we demonstrate the feasibility of our
method by applying it to the quark-meson model
\cite{Jungnickel:1995fp,Schaefer:2004en} which serves as a low-energy
effective model for QCD. We present results for the mesonic spectral  
functions in different regimes of the corresponding phase diagram, in
particular around the critical endpoint.

\clearpage

\section{Theoretical Setup}

The FRG represents a powerful tool for non-perturbative calculations in QFT and 
statistical physics. 
It involves introducing an infrared (IR) regulator $R_k$ to suppress
fluctuations  from momentum modes with momenta below the associated
renormalization group (RG) scale $k$. 
Vacuum and thermal fluctuations are then taken into account by removing the
regulator and lowering the scale $k$ from the ultraviolet (UV) cutoff $\Lambda$ 
down to zero.  
The scale dependence of the effective average action $\Gamma_k$ is
given by the following, formally exact flow equation
\cite{Wetterich:1992yh, Morris1994}, 
\begin{equation}
\label{eq:wetterich}
\partial_k \Gamma_k=\tfrac{1}{2}\text{STr}\left\{ \partial_k R_k(\Gamma^{(2)}_k+R_k)^{-1}\right\},
\end{equation}
where $\Gamma^{(2)}_k$ denotes the second functional derivative of the
effective average action, and the supertrace includes internal and
space-time indices as well as the  integration over the loop momentum.
We now apply this flow equation to the quark-meson model by using the
following Ansatz for the effective average action in the zeroth order
derivative expansion, where only the effective potential carries a
scale dependence, 
\beq
\Gamma_{k}[\bar{\psi} ,\psi,\phi]= \int\! d^{4}x \,\Big\{
\bar{\psi} \left({\partial}\!\!\!\slash +
h(\sigma+i\,\vec{\tau}\cdot\vec{\pi}\,\gamma_{5}) -\mu \gamma_0 \right)\psi
+\frac{1}{2} (\partial_{\mu}\phi)^{2}+U_{k}(\phi^2) - c \sigma 
\Big\}\pkt
\label{eq:QM}
\eeq
with $\phi_i=(\sigma,\vec \pi)_i$ and $\phi^2=\sigma^2+\vec
\pi^2$. The effective potential $U_{k}$ is chosen to be of the form 
$U_{\Lambda}=m\phi^2+\lambda\phi^4$ at the UV scale which is here chosen to
be $\Lambda=1$~GeV.  
The four parameters $h$, $m$, $\lambda$ and $c$ are adjusted to reproduce 
physical values for the pion decay constant and the mesonic screening
masses for
$k\to 0$ in the vacuum, i.e.\ $\sigma_0\equiv f_\pi = 93.5$~MeV, 
$m_\pi=138$~MeV and $m_\sigma=509$~MeV, with a constituent quark mass
 $m_\psi$ of roughly $ 300$~MeV.

\begin{figure}[t]
\centering\includegraphics[width=0.9\columnwidth]{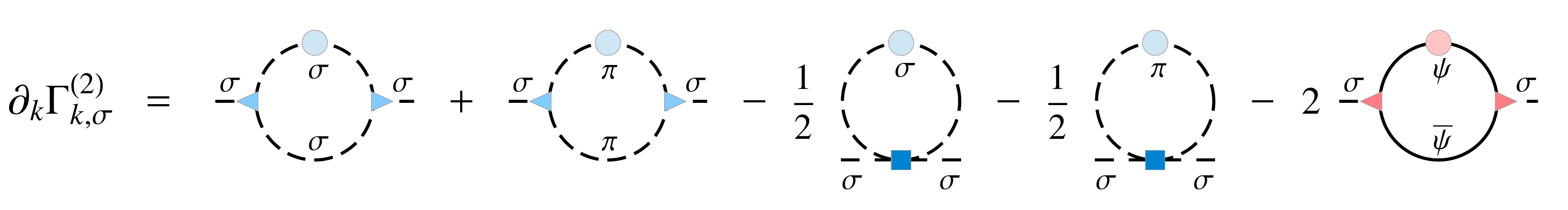}\\
\centering\includegraphics[width=0.9\columnwidth]{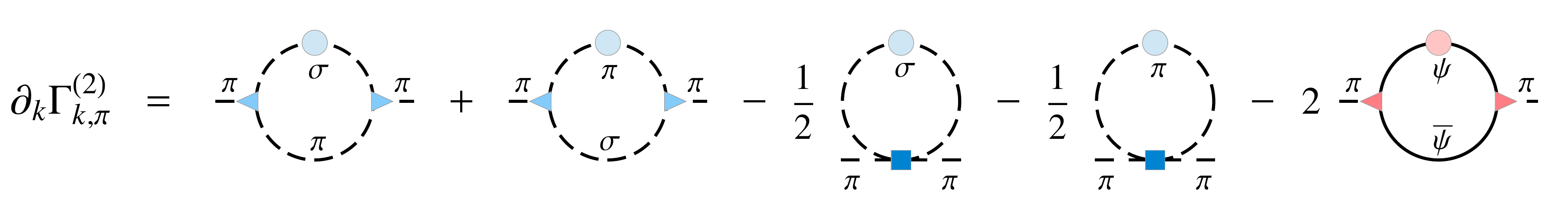}
\caption{(color online) Diagrammatic representation of 
the flow equations for the sigma and pion 2-point functions for the quark-meson model.}
\label{fig:flow_Gamma2} 
\end{figure}

In order to calculate spectral functions we first have to derive flow
equations for  
the 2-point functions.
These are obtained by taking two functional derivatives of Eq. (\ref{eq:wetterich}) 
and are represented diagrammatically in
Fig.~\ref{fig:flow_Gamma2}.\footnote{For explicit expressions for the
  flows of effective potential and 2-point functions, and details 
  on their numerical implementation, see \cite{Tripolt2014}.}
Therein, the quark-meson 3-point vertices are taken to be momentum and
scale-independent, 
$\Gamma^{(2,1)}_{\bar\psi\psi\sigma}=h$ and $\Gamma^{(2,1)}_{\bar\psi\psi\vec\pi}=ih\gamma^5\vec\tau$, 
while the mesonic vertices are scale-dependent as they are extracted from
the solution for the scale-dependent effective potential.
The flow equations for the retarded 2-point functions are then obtained by
analytic continuation using the following two-step procedure: 

\begin{enumerate}
\item We first exploit the periodicity in the discrete external Euclidean
  $p_0=i\,2\pi n T$ of the bosonic and fermionic occupation numbers in
  the flow equations: 
\beq
n_{B,F}(E+i p_0) =  n_{B,F}(E)\pkt
\eeq
\item In a second step, we replace the discrete imaginary external
  energy by a continuous real energy:
\beq
\Gamma^{(2),R}(\omega)=-\lim_{\epsilon\to 0} \Gamma^{(2),E}
(p_0=i\omega-\epsilon).
\eeq
\end{enumerate}
The resulting flow equations for the retarded 2-point functions are
then solved numerically 
with the spectral functions given by 
\beq
\rho(\omega)=\frac{1}{\pi}\frac{\text{Im}\,\Gamma^{(2),R}(\omega)}{\left(\text{Re}\,
\Gamma^{(2),R}(\omega)\right)^2+\left(\text{Im}\,\Gamma^{(2),R}(\omega)\right)^2}.
\eeq

\section{Results}

\begin{figure*}
\includegraphics[width=0.45\columnwidth]{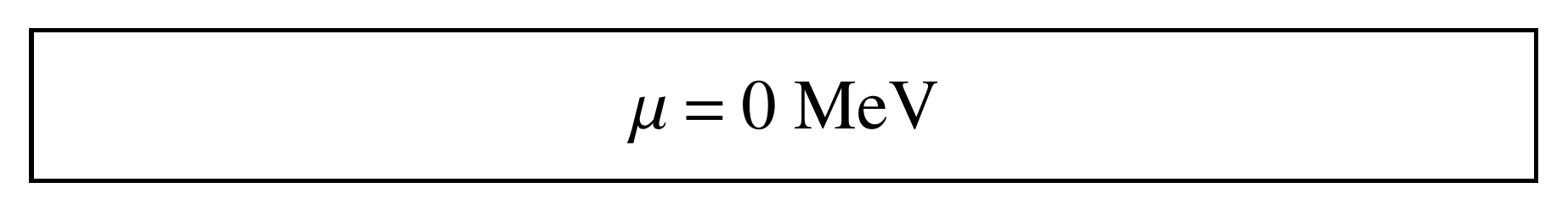}\hspace*{9mm}\includegraphics[width=0.45\columnwidth]{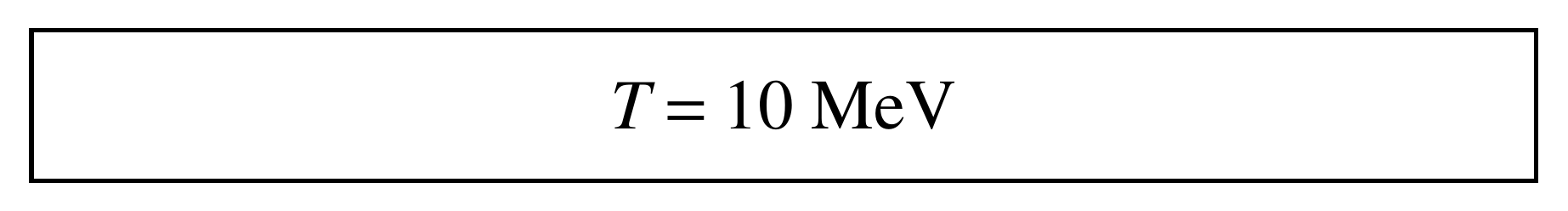}\vspace{1mm}
\includegraphics[width=0.48\columnwidth]{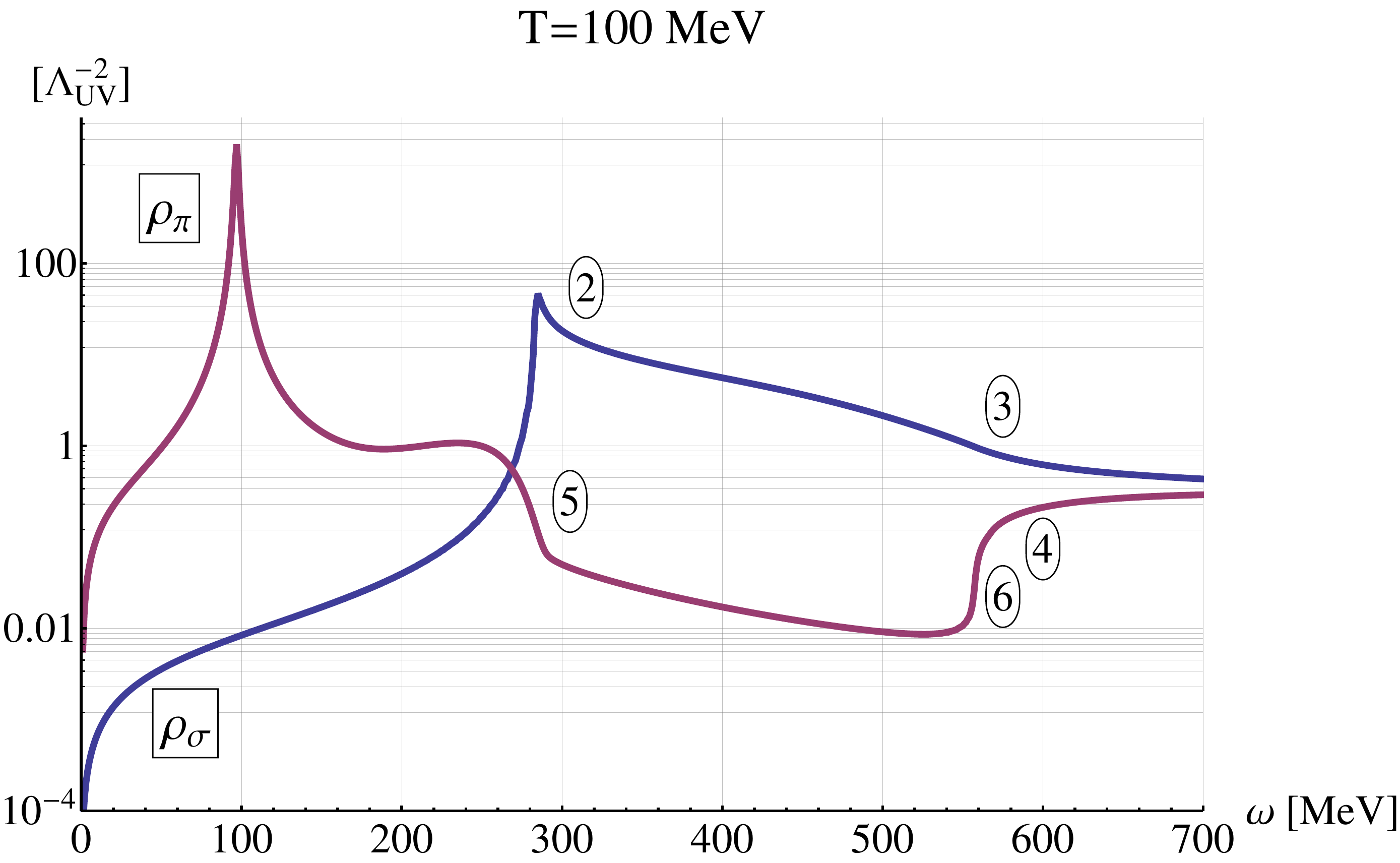}\hspace{5mm}
\includegraphics[width=0.48\columnwidth]{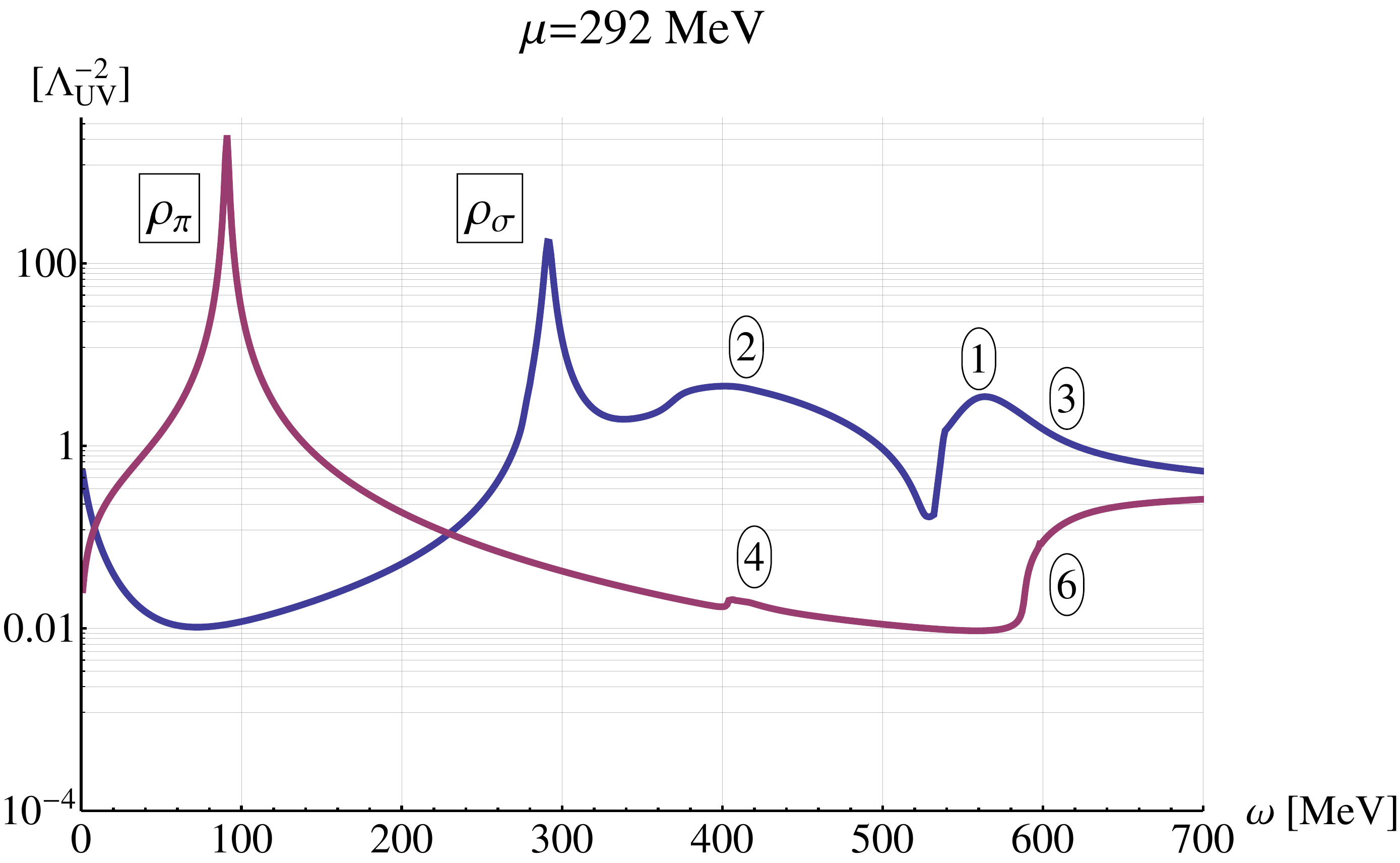}\vspace{3mm}
\includegraphics[width=0.2\columnwidth]{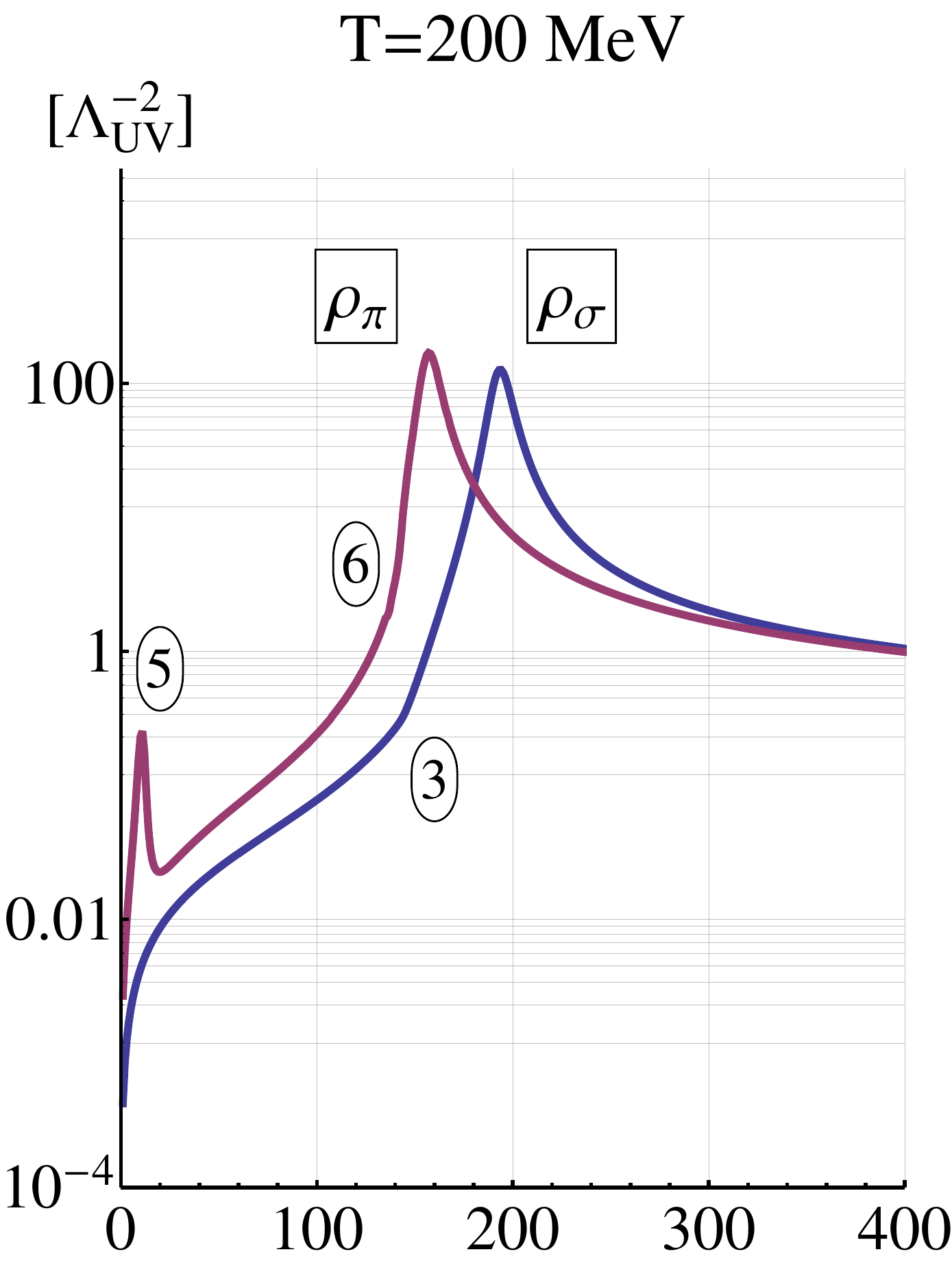}
\includegraphics[width=0.25\columnwidth]{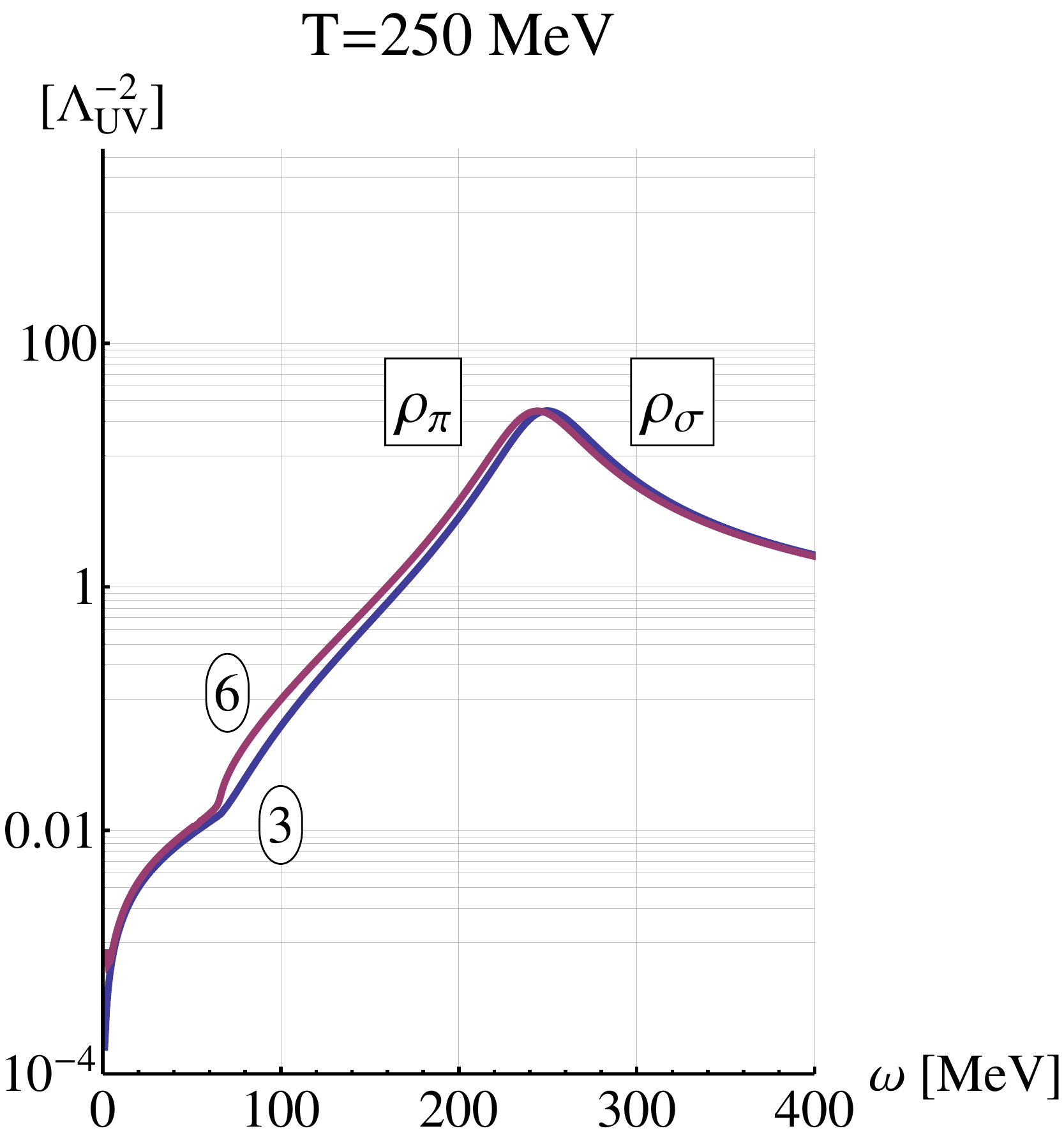}\hspace{11mm}
\includegraphics[width=0.2\columnwidth]{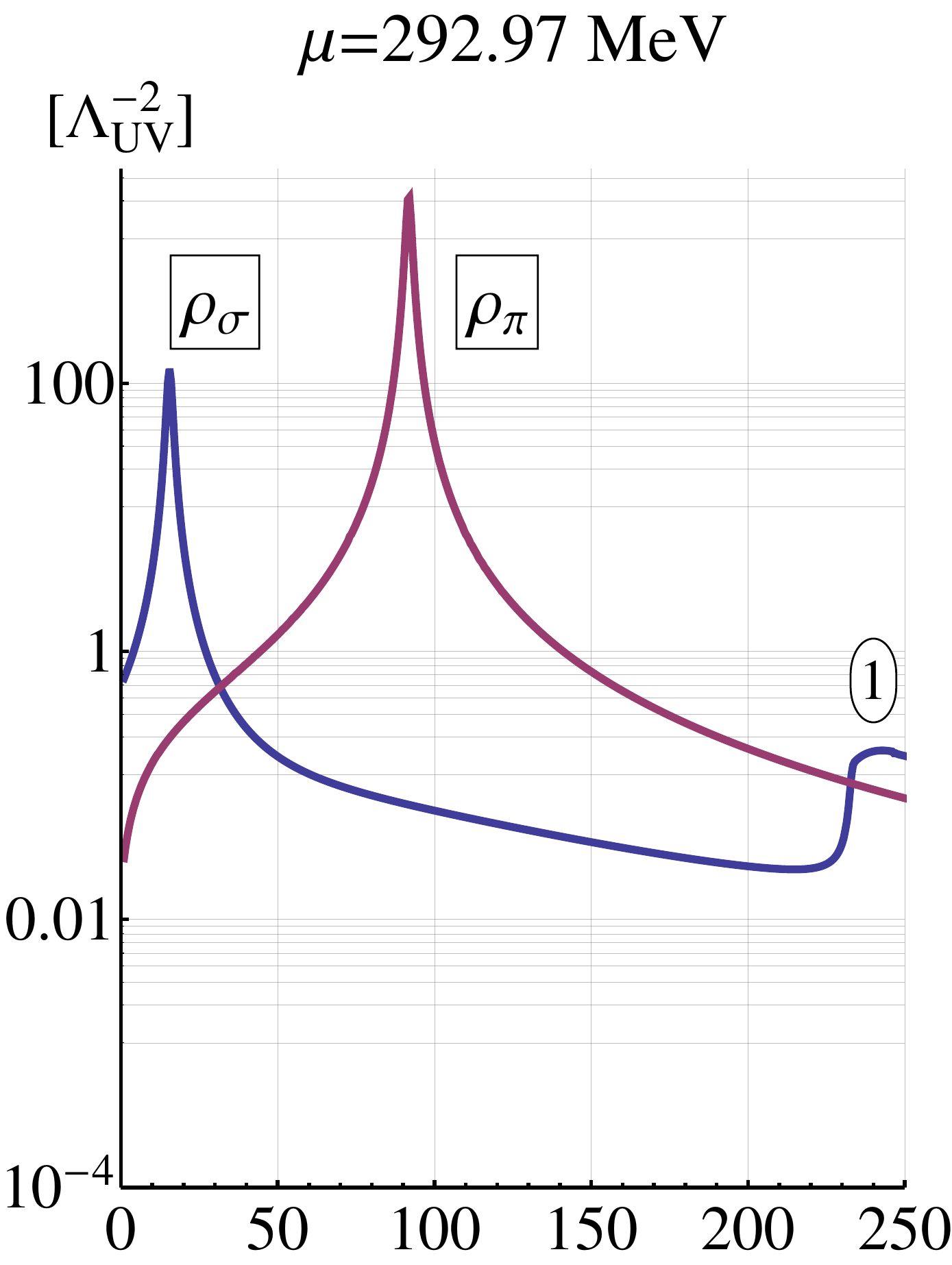}
\includegraphics[width=0.25\columnwidth]{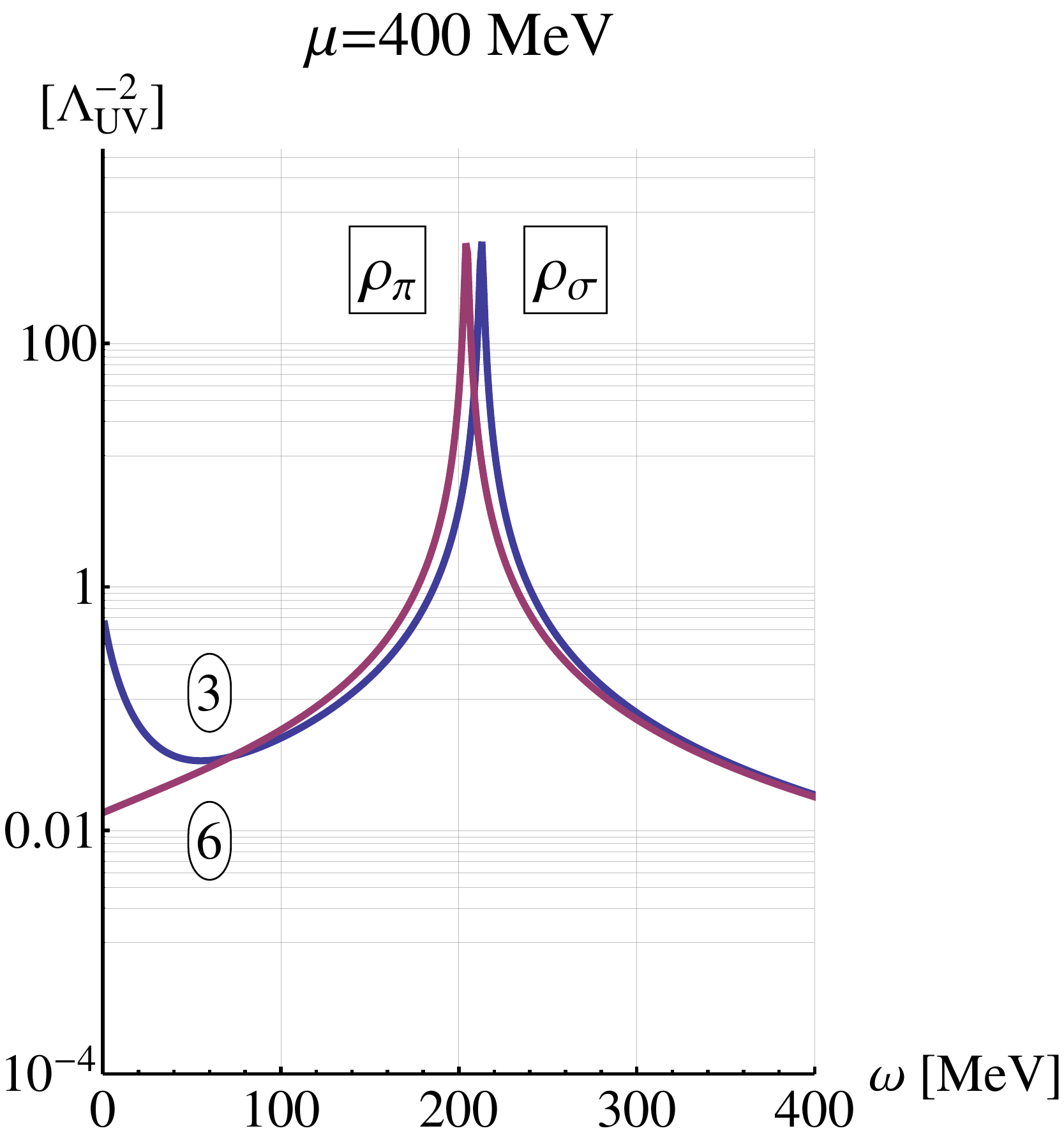}
\caption{(color online) Sigma and pion spectral function from \cite{Tripolt2014}
are shown versus external energy $\omega$ at $\mu=0\,{\rm MeV}$ 
but different temperatures (left column) and at $T=10\,{\rm MeV}$ but different chemical potentials (right column).
Inserted numbers refer to the different processes affecting the spectral functions at corresponding energies. 
1:~$\sigma^* \rightarrow \sigma \sigma$, 
2:~$\sigma^* \rightarrow \pi \pi$, 
3:~$\sigma^* \rightarrow \bar{\psi}\psi$, 
4:~$\pi^* \rightarrow \sigma \pi$, 
5:~$\pi^*\pi \rightarrow \sigma$, 
6:~$\pi^* \rightarrow \bar{\psi}\psi$. 
See text for details.
}
\label{fig:spectralfunctions} 
\end{figure*}

On the left-hand side of Fig.~\ref{fig:spectralfunctions}, the sigma and pion spectral functions, 
$\rho_\sigma(\omega)$ and $\rho_\pi(\omega)$, are shown as a function of the external energy $\omega$ 
at $\mu=0$ and different temperatures. 
Up to $T=100\,{\rm MeV}$ the spectral functions closely 
resemble those in the vacuum, except for an increasing bump due to the
thermal scattering process ${\pi^*\pi\rightarrow \sigma}$
which can only occur at finite temperature, denoted by number 5.
The dominant peak in the pion spectral function signals a stable pion while the sigma spectral function
only shows a broad maximum due to the opening of the ${\sigma^*\rightarrow \pi\pi}$ decay channel for
$\omega\geq 2\,m_\pi$. At higher energies, i.e.\ for $\omega\geq 2\,m_\psi$, both the pion and the
sigma meson may decay into a quark-antiquark pair, giving rise to an increase in the spectral functions.
At $T = 200 \,{\rm MeV}$ and $T = 250 \,{\rm MeV}$, i.e. beyond the chiral crossover transition, 
we observe that the sigma and pion spectral functions become degenerate as expected from the progressing 
restoration of chiral symmetry. 

The right-hand side of Fig.~\ref{fig:spectralfunctions} shows the sigma and pion spectral functions at 
a fixed temperature of $T=10\,{\rm MeV}$ but different values of the quark chemical potential.
For a large range of chemical potentials the spectral functions closely resemble their vacuum structure,
as expected from the Silver Blaze property \cite{Cohen2003}. Only very close to the critical endpoint, which is located at
$\mu= 293\,{\rm MeV}$ and $T=10\,{\rm MeV}$ for our choice of parameters, one observes changes in the sigma
spectral function. At $\mu= 292\,{\rm MeV}$ the sigma meson has already become light enough to be stable, 
giving rise to a peak in the sigma spectral function near $\omega= 300\,{\rm MeV}$. Even closer to the critical
endpoint, i.e.\ at $\mu= 292.97\,{\rm MeV}$ the sigma pole mass has
moved close to zero, indicative of the critical fluctuations near a
second order phase transition. At higher chemical potentials, e.g.\ at
$\mu= 400\,{\rm MeV}$, we again observe 
a degeneration of the spectral functions due to the restoration of chiral symmetry.

\clearpage

\section{Summary and Outlook}

We have presented a new method to obtain real-time quantities like spectral functions 
from the non-perturbative FRG approach. The method is based on an analytic continuation 
from imaginary to real frequencies on the level of the flow equations for the 2-point
functions. It is symmetry preserving and thermodynamically consistent which allows to 
study phase transitions and critical behavior. 
Apart from extensions to other spectral functions, e.g. of the $\rho$ and $a_1$ meson, 
which are of relevance for electromagnetic probes in heavy-ion collisions, there is also the 
exciting possibility to study finite external spatial momenta which will allow for the
computation of transport coefficients.

\section*{Acknowledgements}

This work was supported by the Helmholtz International Center for FAIR within the LOEWE 
initiative of the state of Hesse. R.-A.~T. is furthermore supported by the Helmholtz Research School 
for Quark Matter studies, H-QM, and N.~S. is supported by Grant No. ERC-AdG-290623.


\bibliography{qcd}

\end{document}